# Neural networks approach for mammography diagnosis using wavelets features

Essam A. Rashed [(a),*] and Mohamed G. Awad [(b)]

[(a)] Mathematics Dept., Faculty of Science, Suez Canal Univ., Ismailia 41522, Egypt.
[(b)] Mathematics Dept., Faculty of Education, Suez Canal Univ., Al-Arish, Egypt.

**ABSTRACT**

A supervised diagnosis system for digital mammogram is developed. The diagnosis processes are done by transforming the data of the images into a feature vector using wavelets multilevel decomposition. This vector is used as the feature tailored toward separating different mammogram classes. The suggested model consists of artificial neural networks designed for classifying mammograms according to tumor type and risk level. The results are enhanced from our previous study by extracting feature vectors using multilevel decompositions instead of one level of decomposition. Radiologist-labeled images were used to evaluate the diagnosis system. The results are very promising and show possible guide for future work.

## I. INTRODUCTION

Digital mammography is currently one of the most important tools used for early detection of breast cancer. To achieve the best performance of using digitized mammograms in cancer diagnosis, two main problems may be investigated. The first problem is the classification of cancerous lesions. The second one is to detect the risk level (i.e. benign or malignant). Breast cancer is one of the most dangerous types of cancer among women around the globe. Detecting the cancer in its early steps of creation increases the rate of possible successive treatment [1]. Although the most accurate method in the medical environment is an aggressive biopsy; it still difficult solution according to some risks like discomfort of a patient and high costs. Moreover, the high percentage of negative cases, which is rated 70% to 90% of breast biopsies performed in women, necessitates the search of cheaper and less invasive detection tool [2]. As a result digital mammography have been used in attempts to reduce the negative biopsy ratio and the cost by improving feature analysis and refining criteria for recommendation for biopsy [3-4].

Among all medical imaging techniques that used in cancer diagnosis, digital mammogram is a convenient and easy tool in classifying tumors and many applications in the literature prove their effectiveness in breast cancer diagnosis. Chitre *et al.* used texture measures for the classification of microcalcification regions on effectiveness in breast cancer diagnosis [5]. Another approach is presented by Polakowski *et al.* by developing a model-based vision algorithm using Difference-of-Gaussian (DOG) filters to detect masses and computed nine features. A multilayer perceptron (MLP) neural network was used for the classification of breast masses as benign or malignant. With a dataset of 36 malignant and 53 benign cases, they reported a detection sensitivity of 92% for identifying malignant masses, with 1.8 false positives per image [6]. Kinoshita *et al.* used a combination of shape and texture features. Using a three-layer MLP, they reported 81% accuracy in the classification of benign and malignant breast lesions with a dataset of 38 malignant and 54 benign lesions [7]. In a previous study we developed a CAD system based of selecting wavelets-based features from mammograms and classifying based on MLP. In the proposed system we split the classification steps into four levels which reduces the number of objects while moving from one level to another (e.g. if the mammogram classified as "cancer free" in the first level it is automatically removed from the dataset in the next level of classification). In that study we achieve an interesting successful classification rates which encourage us to enhance it here [8].

In another study we propose an approach for mammograms diagnosis based on selecting a fraction amount of wavelets coefficients with a multilevel of decomposition [9]. Results are enhanced compared with another corresponding method based on single level of wavelets decomposition [10]. We believe that this enhancement is a result of using multilevel of decomposition instead of using one level. The idea of using multilevel wavelets decomposition in mammogram diagnosis is previously used in the literature and proved to be a good choice of mammography features [11]. In this paper a feature vectors is extracted from mammograms based on multilevel wavelets decomposition. These vectors are used to train a MLP for diagnosis mammograms.

[*] Corresponding Author, E-mail: erashed@science.suez.edu.eg

The accuracy of this proposed system is measured using two parameters: Specificity and Sensitivity, which can measured using equation given bellow:

$$Specificty = \frac{TN}{TN+FP}, Sensitivity = \frac{TP}{TP+FN} \quad (1)$$

While, TP is the rate of true positive, TN is the rate of true negative, FP is the rate of false positive, and FN is the rate of false negative [12]. Achieving high specificity means that few cases will be unnecessary recommended for biopsy. While a high sensitivity means that few cancers will be missed. The most important parameter is sensitivity since errors in recognizing cancerous lesions are life-threatening. Errors in recognizing TN's are not life-threatening but they do cause stress and anxiety and waste of resources and may be money.

## II. WAVELETS ANALYSIS

The basic idea of the discrete wavelet transform (DWT), which is detailed in [13], is approximating a signal through a set of basic mathematical functions. The continuous wavelets transform (CWT) of a function $f$ using a wavelet function basis is defined as:

$$f(a,b) = \int f(x)\psi_{a,b}(x)dx \quad (2)$$

While $\psi(x)$ is the mother wavelet function. The basis of wavelet function is obtained by scaling and shifting a signal mother wavelet function.

$$\psi_{a,b}(x) = \frac{1}{\sqrt{a}}\psi(\frac{x-b}{a}) \quad ; \quad a > 0 \quad (3)$$

Where $a$ is the scale factor and $b$ is the shift value. The mother wavelet should only satisfy the zero average condition (i.e. $\int \psi(x)dx = 0$). The DWT is obtained by taking $a = 2$ and $b \in Z$. In the case of 2D signal (i.e. images), the 2D analysis can be performed as a product of two 1D basis functions as shown in the following equation:

$$\psi_{a1,b1,a2,b2}(x_1, x_2) = \psi_{a1,b1}(x_1).\psi_{a2,b2}(x_2) \quad (4)$$

This yield a multiresolution decomposition of the signal into four subbands called the approximation (low frequency component) and details (high frequency component). The approximation **A** is a low resolution of the original image. The details are the coefficients that are neglected during approximation for the horizontal **H**, vertical **V** and diagonal direction **D**. The decomposition process can be iterated with successive approximation being decomposed in turn (multilevel decomposition) as shown in Figure 1. While using wavelets decomposition, important information that represents the structure of original data can be captured. Also, wavelets can capture both texture and information efficiently. In literature, DWT is used effectively in image features extractions [8-12, 14].

In the present study, DWT is used to extract the mammogram feature vectors by applying multilevel of decompositions which reduce the number of values used as a classifier input and at the same time keeps the main features of the image details. Two different mother functions from Daubechies family [13] are used in the multiresolution analysis.

## III. ARTIFICIAL NEURAL NETWORKS

Artificial Neural Networks (ANN) is proved to be effective in medical image processing [15]. In literature, ANN is used to detect cancerous lesions from digital mammography [16-18]. In this paper, a set of ANN are used to classify the mammograms in different levels of detection. In the training phase, wavelets coefficients are used as a network input pattern. The training process continues until a satisfactory classification rate is obtained. In the test phase, the module uses a set of mammogram images to test the system diagnosis evaluation. The number of the output nodes in the neural network module depends on the classification levels of the system. A suit of MLP is used with the classic Back-Propagation learning algorithm [19].

This set of ANN is trained using different values of parameters: number of training steps, stopping criterion, number of hidden neurons, and momentum constant. The output of the network in training phase are set to be binary digit either "0" or "1". The output of the test phase is changed into binary form using the following equation:

$$x_i = \begin{cases} 1 & if \quad x_i = \max(Y) \\ 0 & otherwise \end{cases} \quad (5)$$

Where $Y$ the original is output vector and $x_i$ is the i$^{th}$ element of $Y$.

The classifications levels in Figure 2, consists of three levels, level one detect whether the mammogram is cancerous or cancer free, level two detect the cancer lesion, and level three find the risk level of tumors.

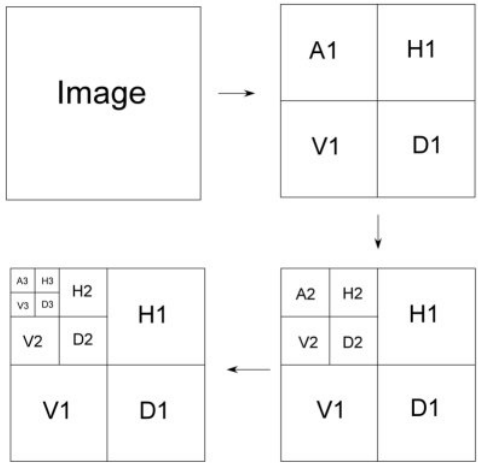 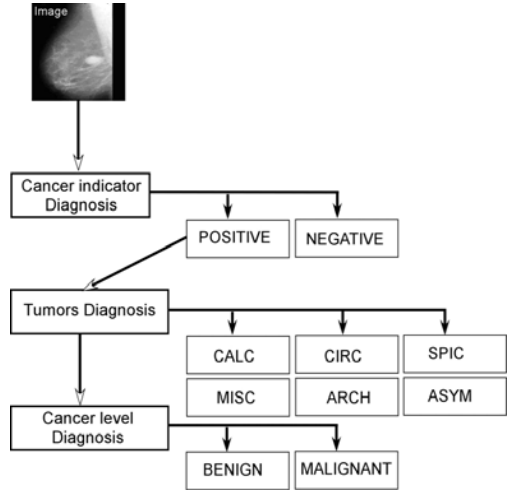

Figure 1: DWT multilevel decomposition of an image    Figure 2: Levels of classifications

## IV. DIAGNOSIS SYSTEM

In the experiment, a set of images from the dataset of the Mammographic Image Analysis Society (MIAS) [20] are used. This dataset were chosen because it covers a wide verity of different cases that covers different types of abnormalities like architectural distortion, asymmetry, microcalcification, circumscribed masses, ill-defined masses, and spiculated lesions. Also, cases cover benign and malignant levels of cancerous tumors. Table 1 summarizes the set of the images in the MIAS database that used in this study. Every different abnormality indicator has different features in shape, brightness, size, and distribution. These features are discussed in details in previous studies. A complete physical features analysis is covered in [11]. Each image is accompanied with an expert radiologist diagnosis proved by a biopsy.

Proposed diagnosis system as shown in Figure 3 which consists of three levels of classification is based on two phases; learning phase and testing phase. In the learning phase three ANN's are trained using feature vectors extracted from wavelets coefficients. The target is based on expert radiologist diagnosis data. In test phase, the previously trained ANN's are used to diagnosis a query mammogram. A set of 330 Regions of Interest (ROI's) of size (128×128 pixels) are extracted from MIAS images which is originally (1024×1024). These ROI's are chosen to contain the abnormality centered.

The multiresolution diagnosis system consists of two basic steps, step one is done by applying DWT on the ROI's for three levels of decompositions to extract image feature vectors. Image feature vector is a combination of the biggest 100 coefficients in each level of decomposition. Using this specific amount of wavelets coefficients are proved to be a good choice that represent the mammogram [10]. Then this set of 300 coefficients that creates the feature vector is used to train three sets of ANN's. The first one is used to distinguish between normal and cancerous ROI's, the second one used for detecting the cancerous lesion, while the last one is used to detect the risk level as shown in Figure 2. In practical evaluation of the diagnosis system, a set of 150 ROI's are taken randomly (with condition of covering all different cases) from the MIAS dataset to extract the feature vectors that is used for the learning phase. After that, all the 330 ROI's are used to evaluate the efficiency of the diagnosis system.

## V. RESULTS

Practically, two different types of wavelets mother functions are used (Daubechies-4, and Daubechies-8) to extract the feature vectors. The classification results of the second level of classification are shown in table 2 in measures of sensitivity and specificity, where S is the number of neurons in the hidden layer and MC is the Momentum Constant. The previously achieved results from [8] are listed also to show the effect of using multilevel of wavelets decomposition. Results obtained from the first and third level of classification are not listed here because they are exactly the same with the previous study which indicates that our contribution here is to refine the detection of cancerous lesion without effecting the detecting of risk level (i.e. benign vs. malignant). It is clear from the result table that experiment number 8 reaches the most optimum conditions for this diagnosis system as we achieve a supreme value of sensitivity and specificity.

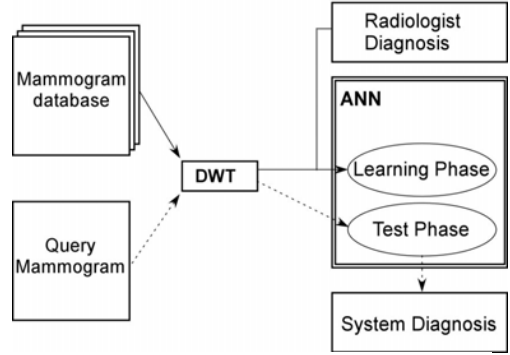

Figure 3: Proposed mammography multiresolution diagnosis system

Table 1: The distribution of cases in MIAS database

| LESION | RISK | # |
|---|---|---|
| Normal (tumors-free) | | 207 |
| Architectural distortion [ARCH] | Benign | 09 |
| | Malignant | 10 |
| Asymmetry [ASYM] | Benign | 06 |
| | Malignant | 06 |
| Microcalcification [CALC] | Benign | 12 |
| | Malignant | 13 |
| Circumscribed masses [CIRC] | Benign | 19 |
| | Malignant | 04 |
| Ill-defined masses [MISC] | Benign | 06 |
| | Malignant | 08 |
| spiculated lesions [SPIC] | Benign | 11 |
| | Malignant | 08 |
| **Total** | | **322** |

Table 2: The diagnosis result

| Exp. # | S | MC | Wavelet | Previous [8] | | Current | |
|---|---|---|---|---|---|---|---|
| | | | | Specificity | Sensitivity | Specificity | Sensitivity |
| 1 | 10 | 0.95 | D-4 | 99.3 | 96.8 | 99.5 | 97.1 |
| 2 | | | D-8 | 98.6 | 98.0 | 98.6 | 98.0 |
| 3 | 10 | 0.95 | D-4 | 99.6 | 99.0 | 99.7 | 99.0 |
| 4 | | | D-8 | 100 | 98.3 | 100 | 98.6 |
| 5 | 10 | 0.8 | D-4 | 99.0 | 99.3 | 99.1 | 99.4 |
| 6 | | | D-8 | 100 | 98.7 | 100 | 98.8 |
| 7 | 20 | 0.95 | D-4 | 100 | 98.7 | 100 | 98.8 |
| 8 | | | D-8 | 100 | 99.0 | 100 | 99.6 |
| 9 | 20 | 0.95 | D-4 | 100 | 99.0 | 100 | 99.0 |
| 10 | | | D-8 | 100 | 99.6 | 100 | 99.6 |
| 11 | 30 | 0.95 | D-4 | 100 | 99.0 | 100 | 99.1 |
| 12 | | | D-8 | 99.6 | 99.3 | 99.7 | 99.6 |
| 13 | 40 | 0.8 | D-4 | 99.6 | 98.7 | 99.8 | 98.8 |
| 14 | | | D-8 | 99.3 | 98.6 | 99.4 | 98.8 |
| 15 | 10 | 0.99 | D-4 | 100 | 98.7 | 100 | 98.6 |
| 16 | | | D-8 | 99.6 | 97.4 | 99.7 | 97.5 |

## VI. CONCLUSION

Using the multilevel decomposition of DWT to extract the features of the digital mammograms is a promising technique to extract features for diagnosis purpose. The proposed diagnosis system achieves good results in classifying the mammograms. Future work could focus on changing the size of features vector and improving the effectiveness the classifier. The presented results in the previous section appear as a positive achievement compared with previous results in the literature. The shortage of this technique is that it needs a supervised classifier with huge database to achieve these high percentage results which is not practically available all the time. Also, this system is built on a single dataset (MIAS) with specific image quality factors; the portability of this system using different mammograms is not tested.

## ACKNOWLDGMENT

The authors would like to thank the CSCBC2006 committee for their support and the anonymous reviewers for their valuable comments.